\documentclass{elsart}
\usepackage{graphicx,amssymb}

\begin{document}
\begin{frontmatter}
\title{Energy, momentum and angular momentum radiation
from chiral cosmic string loops}
\author[inr]{E. Babichev\corauthref{cor}},
\corauth[cor]{Corresponding author.}
\ead{babichev@npd.inr.ac.ru}
\author[inr]{V. Dokuchaev}
\address[inr]{Institute For Nuclear Research of the Russian Academy of
Sciences,
 60th October Anniversary Prospect 7a, 117312 Moscow, Russia}

\begin{abstract}
We derived expressions for energy, momentum and angular momentum losses
due to gravitational and electromagnetic radiation from the closed
superconducting chiral cosmic strings of arbitrary form. The expressions
for corresponding radiation rates into the unit solid angle have the form
of four-dimensional integrals. In the special case of piece-wise linear
strings these formulas are reduced to sums over the kinks. We calculate
numerically the total radiation rates for three examples of string loops
in dependence of current along the string.
\end{abstract}
\end{frontmatter}

\section{Introduction}
We study the gravitational and electromagnetic radiation of energy,
momentum and angular momentum of superconducting closed cosmic strings
with chiral current. Cosmic strings are linear topological defects, that
may have been created during phase transitions in the early Universe (see
e.~g. reviews in \cite{Vilenkin2,Kibble2}). Oscillating ordinary cosmic
strings (without current) radiate only gravitational waves. The
corresponding energy radiation  was studied in
\cite{Vachaspati1,Burden1,Garfinkle1,Vilenkin5,Allen1,Allen2}. Besides
energy, gravitational waves from strings take also momentum
\cite{Vachaspati1,Allen3,Durrer} and angular momentum \cite{Durrer}. It
was found that the rates (averaged per oscillation period) of energy
$\dot{E}$, momentum $\dot{P}$ and angular momentum $\dot{L}$ losses can
be expressed in the following form:
$\dot{E}^{\rm{gr}}=\Gamma_{E}^{\rm{gr}} G\mu^{2}$,
$\dot{P}^{\rm{gr}}=\Gamma_{P}^{\rm{gr}} G\mu^{2}$, $\dot{L}^{\rm{gr}}=
\Gamma_{L}^{\rm{gr}} \mathcal L G\mu^{2}$, where $\Gamma_{E}^{\rm{gr}}\sim
100$, $\Gamma_{P}^{\rm{gr}}\sim 10$ and $\Gamma_{L}^{\rm{gr}}\sim 10$ are
numerical coefficients, depending on the particular string configuration,
$\mathcal{L}$ is length of the string, $\mu$ is string mass per unit
length and used units $\hbar=c=1$.

Witten \cite{Witten1} showed that in some field theory models the cosmic
strings can carry superconducting current which is coupled to some gauge
field. In the case of electromagnetic gauge field the superconducting
cosmic string loops would radiate not only gravitational waves, but also
electromagnetic ones. Equations of motion of superconducting cosmic
strings can be solved analytically \cite{Carter1,Davis2,Vilenkin3} if (i)
the influence of gauge field on the string motion is negligible and (ii)
the current on the string $j^{\mu}$ is chiral, i.~e. $j^{\mu} j_{\mu} =
0$. The gravitational and electromagnetic energy radiated by the single
cusp on chiral cosmic string was studied by Blanco-Pillado and Olum
\cite{Blanco} in the case of small current. The opposite case for current
which is close to maximum value was considered in \cite{Babichev}. If the
string carries the current, then the coefficients $\Gamma_{E}^{\rm{gr}}$,
$\Gamma_{P}^{\rm{gr}}$ and $\Gamma_{L}^{\rm{gr}}$ which determine the
gravitational radiation depend on the current along the string.
Corresponding expressions for electromagnetic radiation have the similar
form: $\dot{E}^{\rm{em}}=\Gamma_{E}^{\rm{em}} \mu q^{2}$,
$\dot{P}^{\rm{em}}=\Gamma_{P}^{\rm{em}} \mu q^{2}$,
$\dot{L}^{\rm{em}}=\Gamma_{L}^{\rm{em}} \mathcal{L} \mu q^{2}$, where
numerical coefficients $\Gamma_{E}^{\rm{em}}$, $\Gamma_{P}^{\rm{em}}$ and
$\Gamma_{L}^{\rm{em}}$ also depend on the current on the string.

In this paper we present the results for gravitational and
electromagnetic radiation of energy, momentum and angular momentum from
chiral cosmic string loops for any value of superconducting current. Due
to the periodic motion of the cosmic string loops the rates of radiation
losses can be expand in series $\dot{E}=\sum\dot{E}_{l}$,
$\dot{P}=\sum\dot{P}_{l}$, $\dot{L}=\sum\dot{L}_{l}$. Here $\dot{E}_{l}$,
$\dot{P}_{l}$ and $\dot{L}_{l}$ are correspondingly the energy, momentum
and angular momentum rates in the $l$-th radiation mode. Usually the
total rates per unit time (averaged over the period) are calculated by
summing of losses in different modes. In practical numerical calculations
the values of $\dot{E}$, $\dot{P}$ and $\dot{L}$ are determined with the
accuracy up to the $l$ of a few hundred. Such calculations may be not
correct because of the slow convergence of the corresponding sums over
$l$ as was pointed out by Allen et al. \cite{Allen1}. See however
\cite{Allen1,Allen2,Allen3} where for some special cases of ordinary
string loops the summation over $l$ was done and analytic expressions for
the total energy and momentum rates into gravitational waves were
obtained. We perform in the following the summation over radiation modes
analytically and derive the formulas for energy, momentum and angular
momentum rates into the both the gravitational and electromagnetic
radiation from chiral string loops of general configuration. The
corresponding radiation rates into the unit solid angle are reduced to the
four-dimensional integrals which in general case can be calculated only
numerically. For chiral piece-wise linear loops these formulas lead to
analytic expressions for the energy, momentum and angular momentum
radiation into the unit solid angle.

We considered three examples of chiral string loops and calculated the
total radiated energy, moment and angular moment per unit time in
dependence of current on the string. The first and the second examples
are piece-wise linear loops (or "kinky" loops), and the third example is
a hybrid of piece-wise loop and smooth loop (namely, the $a$-loop is
smooth and $b$-loop is piece-wise loop). Unfortunately we are unable to
present the results for kinkless cosmic loops (i.~e. for smooth $a$ and
$b$-loops), because it would take an enormous amount of computer time.

The paper is organized as follows. In Section~\ref{sec:11} we review some
general properties of chiral cosmic strings. In Section~\ref{sec:gr} we
derive new expressions for gravitational radiation rates of energy,
momentum and angular momentum by chiral loops of general configuration
into the unit solid angle. These expressions are reduced to the
four-dimensional integrals where the summation over all radiation modes
were performed analytically. In Section~\ref{sec:el} we derive the similar
formulas for electromagnetic radiation rates. In Section~\ref{sec:ex} we
present numerical calculations of electromagnetic and gravitational
radiation rates for some illustrative examples of chiral loops and study
the properties of chiral strings radiation in dependence of current. In
conclusion Section~\ref{sec:co} we describe the obtained results and
discuss some qualitative features of gravitational and electromagnetic
radiation from chiral loops.

\section{Chiral string motion in flat space-time} \label{sec:11}

While moving cosmic string sweeps out a two-dimensional world-sheet in
the Minkowskian space-time. The four-dimensional coordinates of string are
functions of two world-sheet parameters $x^{\mu}=x^{\mu}(\sigma^{a})$,
where indexes $a$ take values $0$, $1$ and $\sigma^{a}$ are
correspondingly the coordinates on a two-dimensional world-sheet. The
convenient gauge choice is such that $\sigma^{0}$ is the Minkowskian time
$t$ and $\sigma^{1}$ parameterizes the string total energy:
\begin{equation}
  \label{sigma}
  E=\mu\int d\sigma.
\end{equation}
In this gauge the general solution of the equations of motion of the
chiral string is \cite{Carter1,Davis2,Vilenkin3}:
\begin{equation}
  \label{ab}
  x^{0}=t, \quad
  \mathbf{x}(t,\sigma)=
  \frac{\mathcal L}{4\pi}\left[\mathbf{a}(\xi)+\mathbf{b}(\eta)\right],
\end{equation}
where ${\mathcal L}$ is the invariant length of the string,
$\mathbf{a}(\xi)$ and $\mathbf{b}(\eta)$ are arbitrary vector functions of
$\xi=(2\pi/{\mathcal L})(\sigma-t)$ and $\eta=(2\pi/{\mathcal
L})(\sigma+t)$ obeying the following conditions:
\begin{equation}
  \label{ab=1k}
  {\mathbf{a}'}^{2}=1,\quad
  {\mathbf{b}'}^{2}=k^{2}(\eta)\leq 1.
\end{equation}
In the case of closed chiral strings (loops) the vector functions
$\mathbf{a}(\xi)$ and $\mathbf{b}(\eta)$ form closed loops, called $a$-
and $b$- loops. The function $k(\eta)$ in (\ref{ab=1k}) may be expressed
as follows \cite{Vilenkin3}:
\begin{equation}
  \label{k}
  k^{2}(\eta)=1-\frac{4F'^{2}(\eta)}{\mu},
\end{equation}
where function $F(\eta)$ defines in turn the auxiliary scalar field
\begin{equation}
  \label{phi}
  \phi(\sigma,t)=\frac{\mathcal L}{2\pi} F(\eta).
\end{equation}
According to (\ref{phi}) the scalar field $\phi(\sigma,t)$ is an
arbitrary function of the only parameter $\eta$. The four-dimensional
current on the string is expressed through this scalar field
$\phi(\sigma,t)$ in the following way \cite{Vilenkin4}:
\begin{equation}
  \label{j}
  j^{\mu}(\mathbf{x},t)=q \int d\sigma
  \phi'(\sigma,t)(x'^{\mu}-\dot{x}^{\mu})
  \delta^{(3)}\left(\mathbf{x}-\mathbf{x}(\sigma,t)\right),
\end{equation}
where $x'$ denotes $\partial x/\partial\sigma$ and $\dot x$ denotes
$\partial x/\partial t$. The energy-momentum tensor of the string in this
gauge is
\begin{equation}
  \label{T}
  T^{\mu\nu}= \mu\int
  d\sigma\left(\dot{x}^{\mu}\dot{x}^{\nu}-x'^{\mu}x'^{\nu}\right)
  \delta^{(3)}\left(\mathbf{x}-\mathbf{x}(\sigma,t)\right).
\end{equation}
Correspondingly the total momentum and angular momentum of the string are
\begin{equation}
  \label{P} \mathbf{P}=\mu \int d \sigma
  \dot{\mathbf{x}}(\sigma,t),
  \end{equation}
\begin{equation}
  \label{L} \mathbf{L}=\mu \int d \sigma \mathbf{x}(\sigma,t)
  \times \dot{\mathbf{x}}(\sigma,t).
\end{equation}

\section{Gravitational radiation from chiral loops} {\label{sec:gr}}

Let us consider the periodic system with period $T$. The Fourier
transform of energy-momentum tensor of this system
$T^{\mu\nu}(\mathbf{x},t)$ can be given by \cite{Durrer}:
\begin{equation}
  \label{Fou:T}
  \hat{T}^{\mu\nu}(\omega_{l},\mathbf{n})=
  \frac{1}{T}\int_0^T dt\int d^{3}x T^{\mu\nu}(\mathbf{x},t)
  e^{i\omega_{l}(t-\mathbf{n}\mathbf{x})},
\end{equation}
where $\omega_{l}=2\pi l/T$ and $\mathbf{n}$ is an arbitrary unit vector.
It is useful to define also the Fourier transform of the first moment:
\begin{equation}
\label{Fou:Tx} \hat{T}^{\mu\nu p}(\omega_{l},\mathbf{n})=\frac{1}{T}\int
dt \int d^{3}x T^{\mu\nu}(\mathbf{x},t) x^{p}
e^{i\omega_{l}(t-\mathbf{n}\mathbf{x})}.
\end{equation}
Let us further define for convenience the four-dimensional symbol
$n^{\mu}\equiv(1,\mathbf{n})$. For any periodic system the corresponding
gravitational energy, momentum and angular momentum radiation rates
(averaged over the period $T$) per solid angle $d\Omega$ is given by
series
\begin{equation}
\label{sum:EPA} \frac{d \dot{P}^{\mu}}{d \Omega}=\sum_{n=1}^{\infty}
\frac{d \dot{{P}^{\mu}} ({\omega}_{n})}{d \Omega},\quad \frac{d
\dot{\mathbf{L}}}{d \Omega}=\sum_{n=1}^{\infty} \frac{d \dot{\mathbf{L}}
({\omega}_{n})}{d \Omega},
\end{equation}
where \cite{Weinberg}
\begin{equation}
  \label{P:ome} \frac{d \dot{P}^{\mu}(\omega)}{d\Omega}=
  -n^{\mu}\frac{G\omega^{2}}{\pi}P_{ij}P_{lm}[\hat{T}^{*}_{il}\hat{T}_{jm}-
  \frac{1}{2}\hat{T}^{*}_{ij}\hat{T}_{lm}]
\end{equation}
and \cite{Durrer}
\begin{eqnarray}
  \label{L:ome}
  &\frac{d \dot{L}_{i}(\omega)}{d\Omega}&=-
 \frac{G}{2\pi}\epsilon^{ijk}n^{j}\left[i\omega
  n^{l}P^{pq}(3\hat{T}^{*}_{kl}\hat{T}_{qp}+6\hat{T}^{*}_{kp}\hat{T}_{ql})
\right.\nonumber\\
  &+& \!\!\!\!\!\!\!\left. \omega^{2}P^{lm}P^{pq}(2\hat{T}^{*}_{kmq}\hat{T}_{lp}-
  2\hat{T}^{*}_{km}\hat{T}_{lpq}-
  \hat{T}^{*}_{lpk}\hat{T}_{mq}+\frac{1}{2}\hat{T}^{*}_{lmk}\hat{T}_{pq})
  +c.c\right].
\end{eqnarray}
Here $P_{ij}=\delta_{ij}-n_{i}n_{j}$ is the projection operator to the
plane perpendicular to the unit vector $\mathbf{n}$. It is possible to
simplify (\ref{P:ome}) and (\ref{L:ome}) further by rewriting them in the
corotating basis $(\mathbf{e}_{1},\mathbf{e}_{2},\mathbf{e}_{3})\equiv
(\mathbf{n},\mathbf{v},\mathbf{w})$, where $\mathbf{v}$ and $\mathbf{w}$
are the arbitrary unit vectors, perpendicular each other and to vector
$\mathbf{n}$. In this corotating basis (\ref{P:ome}) and (\ref{L:ome})
transform to the form \cite{Durrer}:
\begin{equation}
  \label{P:dur} \frac{d \dot{P^{\mu}} ({\omega})}{d
  \Omega}=n^{\mu}\frac{G {\omega}^{2}}{\pi} [{\tau}^{*}_{pq}
  {\tau}_{pq}-\frac{1}{2}{\tau}^{*}_{qq}{\tau}_{pp}],
\end{equation}
\begin{equation}
    \label{L:dur} \frac{d \dot{\mathbf{L}} ({\omega})}{d
    \Omega}=\frac{d \dot{L}_{2}}{d \Omega} \mathbf{v} + \frac{d
    \dot{L}_{3}}{d \Omega} \mathbf{w},
\end{equation}
where
\begin{eqnarray}
  \label{L:dur:v} \frac{d\dot{L}_{2}}{d\Omega}&=&
  \frac{G}{2\pi}[-i\omega(3{\tau}^{*}_{13}{\tau}_{pp}
  +6{\tau}^{*}_{3p}{\tau}_{p1})\nonumber\\
  &-&
  {\omega}^{2}(2{\tau}^{*}_{3pq}{\tau}_{pq}-2{\tau}^{*}_{3p}{\tau}_{pqq}-
  {\tau}^{*}_{pq3}{\tau}_{pq}+\frac{1}{2}{\tau}^{*}_{qq3}{\tau}_{pp})
  +c.c],\nonumber \\
  \frac{d\dot{L}_{3}}{d \Omega}&=&
  \frac{G}{2\pi}[ i\omega(3{\tau}^{*}_{12}{\tau}_{pp}
  +6{\tau}^{*}_{2p}{\tau}_{p1})\nonumber\\
  &+&
  {\omega}^{2}(2{\tau}^{*}_{2pq}{\tau}_{pq}-2{\tau}^{*}_{2p}{\tau}_{pqq}-
  {\tau}^{*}_{pq2}{\tau}_{pq}+\frac{1}{2}{\tau}^{*}_{qq2}{\tau}_{pp})
  +c.c].
\end{eqnarray}
Here  $\tau_{pq}$ and $\tau_{pqr}$ are correspondingly Fourier-transforms
of an energy-momen\-tum tensor and its first moment in new the corotating
basis. Note that only indexes $p,q$ with values $2$ and $3$ appear in the
equations (\ref{P:dur}) and (\ref{L:dur:v}). Fourier-transforms
$\tau_{pq}$ for chiral loops can be expressed in the following way:
\begin{equation}
  \label{tau:IY}
  {\tau}_{pq}(\omega_{l},\mathbf{n})=-\frac{{\mathcal L}\mu}{2}[I_{p}(l)
  Y_{q}(l)+Y_{p}(l) I_{q}(l)],
\end{equation}
where functions $I_{p}(l)$ and $Y_{q}(l)$ are expressed through the
``fundamental integrals'':
\begin{eqnarray}
  \label{IY}
  I_{i}(l)&\equiv&
  \frac{1}{2\pi}\int_{0}^{2\pi} d\xi e^{-il[\xi
  +\mathbf{n}\mathbf{a}(\xi)]}
  \mathbf{a}'(\xi)\mathbf{e_{i}}, \nonumber \\
  Y_{j}(l)&\equiv&
  \frac{1}{2\pi}\int_{0}^{2\pi} d\eta e^{il[\eta
  -\mathbf{n}\mathbf{b}(\eta)]} \mathbf{b}'(\eta)\mathbf{e_{j}}.
\end{eqnarray}
Similarly for the first moment (\ref{Fou:Tx}) one can find:
\begin{eqnarray}
  \label{taux:MN}
  &{\tau}_{ijk}&(\omega_{l},\mathbf{n})=\nonumber\\
  &-&\frac{\mathcal{L}^{2}\mu}{8
  \pi}[I_{i}(l) N_{jk}(l)+I_{j}(l) N_{ik}(l)+Y_{i}(l)
  M_{jk}(l)+Y_{j}(l) M_{ik}(l)],
\end{eqnarray}
where
\begin{eqnarray}
  \label{MN}
  M_{ij}(l)&\equiv&
  \frac{1}{2\pi}\int_{0}^{2\pi} d\xi
  e^{-il[\xi+\mathbf{n}\mathbf{a}(\xi)]}\,(\mathbf{a}'(\xi)\mathbf{e_{i}})\,
  (\mathbf{a}(\xi)\mathbf{e_{j}}),\nonumber\\
  N_{ij}(l)&\equiv&\frac{1}{2\pi}\int_{0}^{2\pi}
  d\eta e^{il[\eta-
  \mathbf{n}\mathbf{b}(\eta)]}\,(\mathbf{b}'(\eta)\mathbf{e_{i}})\,
  (\mathbf{b}(\eta)\mathbf{e_{j}}).
\end{eqnarray}
The crucial point of our following calculations is the summations over
the all mode numbers $l$ in expressions (\ref{sum:EPA}) for the requested
rates of radiated gravitational energy, momentum and angular momentum. To
do this we first integrate expressions (\ref{IY}) and (\ref{MN}) in parts
to get additional $l$ in the denominator. For example for function
$I_{i}$ we have:
\begin{eqnarray}\label{explan}
  I_{i}(l)&=&\frac{1}{2\pi}\int_{0}^{2\pi} d\xi \left[e^{-il(\xi
  +\mathbf{n}\mathbf{a})}(1+\mathbf{n}\mathbf{a'})\right]
  \frac{\mathbf{a}'\mathbf{e_{i}}}{1+\mathbf{n}\mathbf{a'}}= \nonumber\\
  &-&\frac{1}{2\pi i l}\frac{\mathbf{a}'\mathbf{e_{i}}}{1+\mathbf{n}
  \mathbf{a'}}e^{-il(\xi+\mathbf{n}\mathbf{a})}\Big\vert_{0}^{2\pi}
  +\frac{1}{2\pi i l}\int_{0}^{2\pi} d\xi\,
  \left[\frac{\mathbf{a'}\mathbf{e_{j}}}{1+
  \mathbf{n}\mathbf{a'}}\right]'
  e^{-il(\xi +\mathbf{n}\mathbf{a})}.
\end{eqnarray}
The first term in last expression turns to zero because of the periodicity
of $a$- and $b$- loops.  In a similar way the expressions for functions
$Y_{j}$, $M_{ij}$ and $M_{ij}$ can be integrated by parts. Finally we
obtain:
\begin{eqnarray}\label{IYMN1}
  I_{i}&=&\frac{1}{2\pi i l}\int_{0}^{2\pi} d\xi\,\mathcal{I}_{i}
  e^{-il(\xi +\mathbf{n}\mathbf{a})},\quad Y_{j}=-\frac{1}{2\pi
  i l}\int_{0}^{2\pi} d\eta\,\mathcal{Y}_{j}
  e^{il(\eta-\mathbf{n}\mathbf{b})},\nonumber\\
  M_{ij}&=&\int_{0}^{2\pi} d\xi\left(\frac{1}{2\pi i l}
  \mathcal{M}_{ij}-\frac{1}{2\pi l^{2}}
  \tilde{\mathcal{M}}_{ij}\right)
  e^{-il(\xi +\mathbf{n}\mathbf{a})},\\
  N_{ij}&=&-\int_{0}^{2\pi}d\eta\left(\frac{1}{2\pi i l}
  \mathcal{N}_{ij}+\frac{1}{2\pi l^{2}}
  \tilde{\mathcal{N}}_{ij}\right)
  e^{il(\eta-\mathbf{n}\mathbf{b})},\nonumber
\end{eqnarray}
where
\begin{eqnarray}\label{IYMN2}
  \mathcal{I}_{i}&=&\left[\frac{\mathbf{a'}\mathbf{e_{i}}}{1+
  \mathbf{n}\mathbf{a'}}\right]',\quad
  \mathcal{Y}_{j}=\left[\frac{\mathbf{b'}\mathbf{e_{j}}}
  {1-\mathbf{n}\mathbf{b'}}\right]',\nonumber\\
  \mathcal{M}_{ij}&=&\left[\frac{\mathbf{a'}\mathbf{e_{i}}}
  {1+\mathbf{n}\mathbf{a'}}\right]'(\mathbf{a}\mathbf{e_{j}}),\quad
  \tilde{\mathcal{M}}_{ij}=\left[\frac{(\mathbf{a'}\mathbf{e_{i}})
  (\mathbf{a'}\mathbf{e_{j}})}
  {(1+\mathbf{n}\mathbf{a'})^{2}}\right]',\\
  \mathcal{N}_{ij}&=&\left[\frac{\mathbf{b'}\mathbf{e_{i}}}
  {1-\mathbf{n}\mathbf{b'}}\right]'(\mathbf{b}\mathbf{e_{j}}),\quad
  \tilde{\mathcal{N}}_{ij}=\left[\frac{(\mathbf{b'}\mathbf{e_{i}})
  (\mathbf{b'}\mathbf{e_{j}})}{(1-\mathbf{n}\mathbf{b'})^{2}}\right]'.\nonumber
\end{eqnarray}
Substituting (\ref{IYMN1}) into (\ref{tau:IY}) and
(\ref{taux:MN}) we find:
\begin{eqnarray}\label{tau:IYMN1}
\tau_{ij}&=&-\frac{{\mathcal
L}\mu}{8\pi^{2}l^{2}}\int_{0}^{2\pi}\int_{0}^{2\pi} d\xi d\eta\,
\mathcal{T}_{ij}\, e^{-il[\xi-\eta
+\mathbf{n}(\mathbf{a}(\xi)+\mathbf{b}(\eta))]},\nonumber\\
\tau_{ijk}&=&-\frac{{\mathcal
L^{2}}\mu}{32\pi^{3}l^{2}}\int_{0}^{2\pi}\int_{0}^{2\pi} d\xi
d\eta\, \left(\mathcal{T}_{ijk}+\frac{1}{i l
}\tilde{\mathcal{T}}_{ijk}\right)\, e^{-il[\xi-\eta
+\mathbf{n}(\mathbf{a}(\xi)+\mathbf{b}(\eta))]},
\end{eqnarray}
where
\begin{eqnarray}\label{htau:IY}
\mathcal{T}_{ij}&=&\mathcal{I}_{i} \mathcal{Y}_{j}+
\mathcal{I}_{j} \mathcal{Y}_{i},\nonumber\\
\mathcal{T}_{ijk}&=&\mathcal{I}_{i}\mathcal{N}_{jk}+
\mathcal{I}_{j}\mathcal{N}_{ik} + \mathcal{Y}_{i}\mathcal{M}_{jk}+
\mathcal{Y}_{j}\mathcal{M}_{ik},\\
\tilde{\mathcal{T}}_{ijk}&=&-\mathcal{I}_{i}\tilde{\mathcal{N}}_{jk}-
\mathcal{I}_{j}\tilde{\mathcal{N}}_{ik} +
\mathcal{Y}_{i}\tilde{\mathcal{M}}_{jk}+
\tilde{\mathcal{Y}}_{j}\mathcal{M}_{ik}.\nonumber
\end{eqnarray}
In turn substituting (\ref{tau:IYMN1}) into (\ref{P:dur}) and
(\ref{L:dur:v}) we find the radiation rates of $E$, ${\mathbf P}$ and
${\mathbf L}$ on the particular eigen frequency $\omega_{l}=2\pi l/T$:
\begin{equation}\label{P:my}
  \frac{d\dot{P^{\mu}}({\omega})}{d\Omega}
  =n^{\mu}\frac{G\mu^{2}}{4\pi^{3}l^{2}}\int
  d^{4}\xi\mathcal{P} \cos(l\Delta x),
\end{equation}
\begin{eqnarray}
  \label{L:my} \frac{d\dot{L}_{\rm{v}}}{d \Omega}&=&-
  \frac{G{\mathcal L}\mu^{2}}{16\pi^{4}}\int
  d^{4}\xi\left[\frac{\sin(l\Delta x)}{l^{3}}
  (3\lambda_{2}+\tilde{\Lambda}_{2})+ \frac{\cos(l\Delta
  x)}{l^{2}}\Lambda_{2}\right],\nonumber\\
  \frac{d\dot{L}_{\rm w}}{d \Omega}&=&
  \frac{G{\mathcal L}\mu^{2}}{16\pi^{4}}\int
  d^{4}\xi\left[\frac{\sin(l\Delta x)}{l^{3}}
  (3\lambda_{3}+\tilde{\Lambda}_{3})+
  \frac{\cos(l\Delta x)}{l^{2}}\Lambda_{3}\right],
\end{eqnarray}
here we denoted:
\begin{eqnarray}\label{delx}
  \Delta x&=& \xi-\xi'-(\eta-\eta')+
  \mathbf{n}[\mathbf{a}(\xi)-\mathbf{a}(\xi')+\mathbf{b}(\eta)-
  \mathbf{b}(\eta')],\nonumber\\
  \mathcal{P}&=&\mathcal{T}'_{pq}\mathcal{T}_{pq}-
  \frac{1}{2}\mathcal{T}'_{qq}\mathcal{T}_{pp},\nonumber\\
  \lambda_{2}&=&{\mathcal{T}'}_{13}{\mathcal{T}}_{pp}+
  2{\mathcal{T}'}_{3p}{\mathcal{T}}_{p1},\nonumber\\
  \lambda_{3}&=&{\mathcal{T}'}_{12}{\mathcal{T}}_{pp}+
  2{\mathcal{T}'}_{2p}{\mathcal{T}}_{p1},\\
  \Lambda_{2}&=&2{\mathcal{T}'}_{3pq}{\mathcal{T}}_{pq}-
  2{\mathcal{T}'}_{3p}{\mathcal{T}}_{pqq}
  -{\mathcal{T}'}_{pq3}{\mathcal{T}}_{pq}
  +\frac{1}{2}{\mathcal{T}'}_{qq3}{\tau}_{pp},\nonumber\\
  \tilde{\Lambda}_{2}&=&2\tilde{{\mathcal{T}}'}_{3pq}{\mathcal{T}}_{pq}
  +2{\mathcal{T}'}_{3p}\tilde{{\mathcal{T}}}_{pqq}
  -\tilde{{\mathcal{T}}'}_{pq3}{\mathcal{T}}_{pq}
  +\frac{1}{2}\tilde{{\mathcal{T}}'}_{qq3}{\tau}_{pp},\nonumber\\
  \Lambda_{3}&=&2{\mathcal{T}'}_{2pq}{\mathcal{T}}_{pq}
  -2{\mathcal{T}'}_{2p}{\mathcal{T}}_{pqq}
  -{\mathcal{T}'}_{pq2}{\mathcal{T}}_{pq}
  +\frac{1}{2}{\mathcal{T}'}_{qq2}{\tau}_{pp},\nonumber\\
  \tilde{\Lambda}_{3}&=&2\tilde{{\mathcal{T}}'}_{2pq}{\mathcal{T}}_{pq}
  +2{\mathcal{T}'}_{2p}\tilde{{\mathcal{T}}}_{pqq}
  -\tilde{{\mathcal{T}}'}_{pq2}{\mathcal{T}}_{pq}
  +\frac{1}{2}\tilde{{\mathcal{T}}'}_{qq2}{\tau}_{pp}. \nonumber
\end{eqnarray}
It is assumed that integration in (\ref{P:my}) and (\ref{L:my}) is over
four-dimensional cube with side $(0,2\pi)$ and notation $d^{4}\xi=d\xi\,
d\xi'\, d\eta\, d\eta'$ is introduced. Now we found the desired form of
expressions (\ref{P:my}) and (\ref{L:my}) suitable for making summations
over modes $l$. Using the known values for infinite series
\cite{Gradshteyn}
\begin{eqnarray}
\label{l:sum}
\sum_{l=1}^{\infty}\frac{\cos(lx)}{l^{2}}&=&\frac{1}{4}(x-\pi)^{2}
-\frac{\pi^{2}}{12},\quad0\leq x\leq 2\pi, \nonumber\\
\sum_{l=1}^{\infty}\frac{\sin(lx)}{l^{3}}&=&\frac{1}{12}[(x-\pi)^{3}
-\pi^{2}x]+\frac{\pi^{3}}{12}, \quad 0\leq x\leq 2\pi,
\end{eqnarray}
we obtain from (\ref{P:my}) and (\ref{L:my}) the final expressions for
gravitational radiation of energy, momentum and angular momentum rates:
\begin{equation}\label{P:fin}
  \frac{d\dot{P^{\mu}}}{d\Omega}
  =n^{\mu}\frac{G\mu^{2}}{16\pi^{3}}\int
  d^{4}\xi\mathcal{P}(\Delta
  x\,\rm{mod}\,2\pi-\pi)^{2},
\end{equation}
\begin{eqnarray}
  \label{L:fin} \frac{d\dot{L}_{\rm v}}{d\Omega}&=&
  -\frac{G\mathcal{L}\mu^{2}}{64\pi^{4}}\int
  d^{4}\xi\left\{\left[(\Delta x\,\rm{mod} \,2\pi-\pi)^{3}- \pi^{2}
  \Delta x \,\rm{mod}\,2\pi\right]
  (\lambda_{2}+\frac{1}{3}\tilde{\Lambda}_{2})\right.\nonumber\\
  &&\left. +(\Delta
  x\,\rm{mod}\,2\pi-\pi)^{2}\Lambda_{2}\right\},\\
  \frac{d\dot{L}_{\rm w}}{d\Omega}&=&
  \frac{G\mathcal{L}\mu^{2}}{64\pi^{4}}\int
  d^{4}\xi\left\{\left[(\Delta x\,\rm{mod} \,2\pi-\pi)^{3}- \pi^{2}
  \Delta x \,\rm{mod}\,2\pi\right]
  (\lambda_{3}+\frac{1}{3}\tilde{\Lambda}_{3})\right.\nonumber\\
  &&\left. +(\Delta
  x\,\rm{mod}\,2\pi-\pi)^{2}\Lambda_{3}\right\}.\nonumber
\end{eqnarray}
Note that the integrals in expressions (\ref{P:fin}) and (\ref{L:fin}) do
not contain terms $\pi^{2}/12$ and $\pi^{3}/12$ originated from
(\ref{l:sum}) because of the nullifying of corresponding contributions in
the integrals. The advantage of formulas (\ref{P:fin}) and (\ref{L:fin})
with respect to the corresponding formulas (\ref{P:dur}) and
(\ref{L:dur}) is that there are no summations over modes. Meanwhile due to
the presence of function $\Delta x(\rm{mod})2\pi$ the four-dimensional
integrals in (\ref{P:fin}) and (\ref{L:fin}) can not be reduced into the
product of integrals of smaller dimensions and therefore the numerical
calculations of four-dimensional integrals become more complicated.

\section{Electromagnetic radiation}{\label{sec:el}}

Now let us consider in a similar way the electromagnetic radiation from
any relativistic periodic systems. A retarded solution for electromagnetic
potential $A_{\mu}$ for such a system in Lorentz gauge is given
\begin{equation}
  \label{A:int} A_{\mu}(\mathbf{x},t)=
  -\int \frac{j_{\mu}(\mathbf{x}',t_{ret})}{|\mathbf{x}
  -\mathbf{x}'|}d\mathbf{x}',
\end{equation}
where $j_{\mu}$ is four-dimensional current and we denoted
$t_{ret}=t-|\mathbf{x}-\mathbf{x}'|$. We consider formula (\ref{A:int})
in the limit $r=|\mathbf{x}|\gg |\mathbf{x}'|$. Expanding (\ref{A:int}) in
series on $1/r$ and taking into account the first two terms we obtain:
\begin{equation}
  \label{A:ser} A_{\mu}(\mathbf{x},t)=\frac{1}{r}\int j_{\mu}(\mathbf{x}',t_{ret})d\mathbf{x}' -
  \frac{1}{r^{2}}\int j_{\mu}(\mathbf{x}',t_{ret}) x'^{i}d\mathbf{x}' +O(r^{-3}),
\end{equation}
where  $\mathbf{n}=\mathbf{x}/r$. Then expanding $t_{ret}$ in series on
$|\mathbf{x}'|/r$ we find:
\begin{equation}
  \label{t:ret} t_{ret}=t-r+\mathbf{n}\cdot
  \mathbf{x}'-\frac{1}{2r}P_{ij}x'^{i}x'^{j}+
  O({|\mathbf{x}'|}^{2}/r^{2})|\mathbf{x}'|.
\end{equation}
From (\ref{t:ret}) it follows the useful relation:
\begin{equation}
  \label{A:con} A_{\mu,j}=-A_{\mu,0} n_{j} + O(A_{\mu}/r).
\end{equation}
Similarly to the case of gravitational field ($T^{\mu\nu}\leftrightarrow
j^{\mu}, h^{\mu\nu}\leftrightarrow j^{\mu}$, etc.) we have the Fourier
transforms of current $\tilde{j}^{\mu}$ and its first and second moment
$\tilde{j}^{\mu p}$, $\tilde{j}^{\mu p q}$:
\begin{eqnarray}
   \label{j:Fou}
\tilde{j}^{\mu}(\omega_{l},\mathbf{n})&=&\frac{1}{T}\int_{0}^{T} dt
  \int d^{3}x j^{\mu}(\omega_{l},\mathbf{x})
  e^{i\omega_{l}(t-\mathbf{n}\cdot\mathbf{x})}, \nonumber \\
   \tilde{j}^{\mu p}(\omega_{l},\mathbf{n})&=&\frac{1}{T}\int_{0}^{T} dt\int
   d^{3}x
   j^{\mu}(\omega_{l},\mathbf{x}) x^{p}
   e^{i\omega_{l}(t-\mathbf{n}\cdot\mathbf{x})}, \\
   \tilde{j}^{\mu pq}(\omega_{l},\mathbf{n})&=&\frac{1}{T}\int_{0}^{T} dt\int
   d^{3}x
   j^{\mu}(\omega_{l},\mathbf{x}) x^{p} x^{q}
   e^{i\omega_{l}(t-\mathbf{n}\cdot\mathbf{x})}. \nonumber
\end{eqnarray}
The values (\ref{j:Fou}) obey the following conditions:
\begin{eqnarray}
  \label{3-propj1}
  \tilde{j^{0}}-n^{k}\tilde{j^{k}}&=&0, \nonumber\\
    -i\omega\tilde{j^{0p}}-\tilde{j^{p}}+i\omega
  n_{k}\tilde{j^{kp}}&=&0, \\
  i\omega
  P_{mn}(\,\tilde{j}^{0mn}-n_{p}\tilde{j}^{pmn})+2P_{pq}\tilde{j}^{pq}
  &=&0, \nonumber
\end{eqnarray}
which follow from relations
\begin{eqnarray}
  \label{j:pro}
  j^{\mu}_{,\mu}=0, \nonumber\\
  \int\!
  j^{\mu}(t,\mathbf{x}')_{,\mu}[x'^{p}e^{i\omega(t-
  \mathbf{n}\cdot\mathbf{x}')}]\,dt\,d^{3}x=0,\\
  \int\! j^{\mu}(t,\mathbf{x}')_{,\mu}\left\{[x'^{2}-(\mathbf{n}\cdot
  \mathbf{x}')]
  e^{i\omega(t-\mathbf{n}\cdot\mathbf{x}')}\right\}\,dt\,d^{3}x=0.\nonumber
\end{eqnarray}
Using (\ref{j:Fou}) and (\ref{j:pro}), from (\ref{A:ser}) we obtain:
\begin{eqnarray}
  \label{A} A_{\mu}(\mathbf{x},t)&=&
  \frac{1}{r}\sum_{l=1}^{\infty}e^{-i\omega_{l}(t-r)}
  \left[\tilde{j}_{\mu}(\omega_{l},\mathbf{n}) \!+\!
  \frac{n^{p}}{r}\tilde{j}_{\mu p}(\omega_{l},\mathbf{n}) \!+\!
  \frac{i\omega_{l}}{2r}P^{pq}\tilde{j}_{\mu pq}(\omega_{l},\mathbf{n})\right]
  \nonumber\\
  && + \; c.c.   +O(r^{-3}).
\end{eqnarray}
For calculations of energy and momentum radiation losses we keep in
(\ref{A}) only terms of the order of $1/r$ . The radiation of energy from
the system is determined by the Poynting vector which is equal
\cite{Landau}:
\begin{equation}
  \label{Poi}
  \frac{d \dot{E}^{\rm em}}{d \Omega}=\frac{|\mathbf{E}\times\mathbf{H}|}{4\pi},
\end{equation}
where $\mathbf{E}$ and $\mathbf{H}$ are correspondingly the electric and
magnetic fields. Using (\ref{A}) we obtain from (\ref{Poi}):
\begin{equation}
  \label{Pe:sum}
  \frac{d\dot{P}^{\mu}_{\rm em}}{d\Omega}=
  \sum_{n=1}^{\infty} \frac{d \dot{P^{\mu}}
  ({\omega}_{n})}{d \Omega},
\end{equation}
where
\begin{equation}
  \label{Pe} \frac{d\dot{P}^{\mu}_{\rm em}(\omega)}{d \Omega}
  =n^{\mu}\frac{\omega^{2}}{2\pi}P^{pq}\tilde{j}^{*}_{p} \tilde{j}_{q}.
\end{equation}
Let us calculate now the electromagnetic radiation of angular momentum.
The angular momentum rate per unit solid angle is given \cite{Landau}:
\begin{equation}
  \label{Le0}
  \frac{d\dot{\mathbf{L}}^{\rm em}}{d\Omega}=
  \frac{r^{3}}{4\pi}\left[(\mathbf{n}\!\!\times\!\!
  \mathbf{E})(\mathbf{n}\mathbf{E})+
  (\mathbf{n}\!\!\times\!\!\mathbf{H})(\mathbf{n}
  \mathbf{H})\right].
\end{equation}
In calculations of $(\mathbf{n}\times \mathbf{E})$ and $(\mathbf{n}\times
\mathbf{H})$ it is sufficient to keep only terms of the order of $1/r$.
However the longitudinal components $(\mathbf{n}\mathbf{E})$ and
$(\mathbf{n}\mathbf{H})$ arise from terms of the order of $1/r^{2}$. As a
result in expression (\ref{Le0}) the term $r^{3}$ is canceled. It means
that the distance from the system $r$ is not entered to the final formula
as it should be. Using (\ref{A}) and (\ref{j:pro}) we obtain:
\begin{eqnarray}
  \label{EH}
  (\mathbf{n}\times\mathbf{E})^{i}&=&-\,\epsilon^{ijk}n_{j}A_{k,0}=
  -\sum_{l=1}^{\infty}
  \frac{i\omega_{l}}{r}e^{-i\omega_{l}(t-r)} \epsilon^{ijk}n_{j}\tilde{j}_{k}
  + c.c.,\nonumber\\
  (\mathbf{n}\times\mathbf{H})^{i}&=&
  \sum_{l=1}^{\infty}\frac{i\omega_{l}}{r}e^{-i\omega_{l}(t-r)}
  P^{ij}\tilde{j}_{k} + c.c.,\nonumber\\
  (\mathbf{n}\mathbf{E})&=& -\sum_{l=1}^{\infty}
  \frac{i\omega_{l}}{r^{2}}e^{-i\omega_{l}(t-r)} P^{pq}\tilde{j}_{pq}
  + c.c.,\\
  (\mathbf{n}\mathbf{H})&=&
  \sum_{l=1}^{\infty}\frac{i\omega_{l}}{r^{2}}e^{-i\omega_{l}(t-r)}
  \epsilon^{pqr}n_{p}\tilde{j}_{rq} + c.c.\nonumber
\end{eqnarray}
Substituting (\ref{EH}) in (\ref{Le0}) we find:
\begin{equation}
  \label{Le:sum} \frac{d \dot{\mathbf{L}}^{\rm em}}{d \Omega}=
  \sum_{n=1}^{\infty} \frac{d \dot{\mathbf{L}}^{\rm em}
  ({\omega}_{n})}{d \Omega},
\end{equation}
where
\begin{equation}
  \label{Le} \frac{d \dot{L}_{i}^{\rm em}(\omega)}{d\Omega}=
  \frac{\omega^{2}}{4\pi}
  \left[(\epsilon^{ijk}P_{pq}
  -P_{ik}\epsilon_{jpq})\,n_{j}\,\tilde{j}^{*}_{k}\,
  \tilde{j}_{pq}+\,c.c.\,\right].
\end{equation}
As in the case of gravitational field we will rewrite expressions
(\ref{Pe}) and (\ref{Le}) in corotating basis
$(\mathbf{e}_{1},\mathbf{e}_{2},\mathbf{e}_{3})=
(\mathbf{n},\mathbf{v},\mathbf{w})$:
\begin{equation}
  \label{Pe1} \frac{d\dot{P}^{\mu}_{\rm em}(\omega)}{d
  \Omega}=n^{\mu}\frac{\omega^{2}}{2\pi}\tilde{\iota}^{*}_{p}
  \tilde{\iota}_{p},
\end{equation}
\begin{equation}
  \label{Le1:sum} \frac{d \dot{\mathbf{L}^{\rm em}}
  ({\omega})}{d \Omega}=\frac{d \dot{L}_{2}^{\rm em}}{d \Omega}
  \mathbf{v} + \frac{d \dot{L}_{3}^{\rm em}}{d \Omega} \mathbf{w}.
\end{equation}
Here
\begin{eqnarray}
  \label{Lve} \frac{d \dot{L}^{\rm{em}}_{2}}{d \Omega}&=&-
  \frac{\omega^{2}}{4\pi}[\iota^{*}_{3}\iota_{pp}+
  \iota_{2}^{*}(\iota_{23}-\iota_{32})+c.c.],\nonumber\\
  \frac{d \dot{L}^{\rm{em}}_{3}}{d \Omega}&=&
  \frac{\omega^{2}}{4\pi}[\iota^{*}_{2}\iota_{pp}-
  \iota_{3}^{*}(\iota_{23}-\iota_{32})+c.c.]
\end{eqnarray}
and $\iota^{p},\iota^{pq}$ are components of $j^{\mu}$ and $j^{\mu p}$ in
this corotating basis.

For superconducting chiral strings from expression for current (\ref{j})
we have
\begin{equation} \label{iota}
  {\iota}_{i}(\omega_{l},\mathbf{n})=\frac{\mathcal{L}q
  \sqrt{\mu}}{2}[I_{i}(l)X(l)],
\end{equation}
where the function $I_{i}(l)$ is given by (\ref{IY}) and $X(l)$ is
\begin{equation}
  \label{X}
  X(l)\equiv\frac{1}{2\pi}\int_{0}^{2\pi}d\eta\, e^{il[\eta
  -\mathbf{n}\mathbf{b}(\eta)]} \sqrt{1-|\mathbf{b}'(\eta)|^{2}}.
\end{equation}
Similarly for the first moment $\iota_{pq}$ we obtain
\begin{equation}
  \label{iotax}
  {\iota}_{pq}(\omega_{l},\mathbf{n})=\frac{\mathcal{L}^{2}q\sqrt{\mu}}{8
  \pi}[I_{p}(l) Z_{q}(l)+X(l) M_{pq}(l)],
\end{equation}
where $M_{pq}$ is given by (\ref{MN}) and $Z_{q}$ is
\begin{equation} \label{Z}
  Z_{i}(l)\equiv\frac{1}{2\pi}\int_{0}^{2\pi}\!
  d\eta\,
  e^{il[\eta-\mathbf{n}\mathbf{b}(\eta)]}\sqrt{1-|\mathbf{b}'(\eta)|^{2}}\,
  (\mathbf{b}(\eta)\,\mathbf{e_{i}}).
\end{equation}
We now integrate expressions (\ref{X}) and (\ref{Z}) in parts to get the
additional $l$ in the denominator:
\begin{eqnarray}\label{XZ1}
  X&=&-\frac{1}{2\pi i l}\int_{0}^{2\pi} d\eta\,\mathcal{X}
  e^{il(\eta-\mathbf{n}\mathbf{b})},\nonumber\\
  Z_{j}&=&-\int_{0}^{2\pi}d\eta\left(\frac{1}{2\pi i l}
  \mathcal{Z}_{j}+\frac{1}{2\pi l^{2}}
  \tilde{\mathcal{Z}}_{j}\right)
  e^{il(\eta-\mathbf{n}\mathbf{b})},
\end{eqnarray}
where
\begin{eqnarray}\label{XZ2}
 \mathcal{X}&=&\left[\frac{\sqrt{1-|\mathbf{b'}|^{2}}}
 {1-\mathbf{n}\mathbf{b'}}\right]',\nonumber\\
 \mathcal{Z}_{j}&=&\left[\frac{\sqrt{1-|\mathbf{b'}|^{2}}}
 {1-\mathbf{n}\mathbf{b'}}\right]'(\mathbf{b}\mathbf{e_{j}}),\quad
  \tilde{\mathcal{Z}}_{j}=\left[\frac{\sqrt{1-|\mathbf{b'}|^{2}}
 (\mathbf{b'}\mathbf{e_{j}})}{1-\mathbf{n}\mathbf{b'}}\right]'.
\end{eqnarray}
Substituting (\ref{IY}), (\ref{MN}) and (\ref{XZ1}) into
(\ref{iota}) and (\ref{iotax}) we obtain:
\begin{eqnarray}\label{iota:IXMZ1}
\iota_{i}&=&
\frac{\mathcal{L}q\sqrt{\mu}}{8\pi^{2}l^{2}}\int_{0}^{2\pi}\int_{0}^{2\pi}
d\xi d\eta\, \mathcal{J}_{i}\, e^{-il\{\xi-\eta
+\mathbf{n}[\mathbf{a}(\xi)+\mathbf{b}(\eta)]\}},\nonumber\\
\iota_{ij}&=&
\frac{\mathcal{L}^{2}q\sqrt{\mu}}{32\pi^{3}l^{2}}\int_{0}^{2\pi}\int_{0}^{2\pi}
d\xi d\eta\, \left(\mathcal{J}_{ij}+\frac{1}{i l
}\tilde{\mathcal{J}}_{ij}\right)\, e^{-il\{\xi-\eta
+\mathbf{n}[\mathbf{a}(\xi)+\mathbf{b}(\eta)]\}},
\end{eqnarray}
where
\begin{eqnarray}\label{JJ}
\mathcal{J}_{i}=\mathcal{I}_{i} \mathcal{X},\quad
\mathcal{J}_{ij}=\mathcal{I}_{i}\mathcal{Z}_{j}+\mathcal{X}\mathcal{M}_{ij},\quad
\tilde{\mathcal{J}}_{ij}=-\mathcal{I}_{i}\tilde{\mathcal{Z}}_{j}+\mathcal{X}\tilde{\mathcal{M}}_{ij}.
\end{eqnarray}
Finally, substituting (\ref{iota:IXMZ1}) into (\ref{Pe1}) and (\ref{Lve})
we find the expressions for electromagnetic radiation rates of energy,
momentum and angular momentum into the unit solid angle at frequency
$\omega_{l}$:
\begin{equation}\label{Pe:my}
  \frac{d\dot{P}^{\mu}_{\rm{em}}(\omega_{l})}{d\Omega}
  =n^{\mu}\frac{q^{2}\mu}{8\pi^{3}l^{2}}\int d^{4}\xi\mathcal{P}^{\rm{em}}\cos(l\Delta x),
\end{equation}
\begin{eqnarray}
  \label{Le:my} \frac{d\dot{L}_{\rm v}^{\rm{em}}}{d \Omega}&=&
  \frac{\mathcal{L}q^{2}\mu}{32\pi^{4}}\int
  d^{4}\xi\left[\frac{\sin(l\Delta x)}{l^{3}}
  \tilde{\Lambda}_{2}^{\rm{em}}-\frac{\cos(l\Delta
  x)}{l^{2}}\Lambda_{2}^{\rm{em}}\right],\nonumber\\
  \frac{d\dot{L}_{\rm w}^{\rm{em}}}{d \Omega}&=&-
  \frac{\mathcal{L}q^{2}\mu}{32\pi^{4}}\int
  d^{4}\xi\left[\frac{\sin(l\Delta x)}{l^{3}}
  \tilde{\Lambda}_{3}^{\rm{em}}-
  \frac{\cos(l\Delta x)}{l^{2}}\Lambda_{3}^{\rm{em}}\right],
\end{eqnarray}
where
\begin{eqnarray}\label{hiota:IX}
\mathcal{P}^{\rm{em}}&=&\mathcal{J}'_{p}\mathcal{J}_{p},\nonumber\\
\Lambda_{2}^{\rm{em}}&=&\mathcal{J}'_{3}{\mathcal{J}}_{pp}+
\mathcal{J}'_{2}({\mathcal{J}}_{23}-{\mathcal{J}}_{32}), \quad
\tilde{\Lambda}_{2}^{\rm{em}}=\mathcal{J}'_{3}\tilde{{\mathcal{J}}}_{pp}+
\mathcal{J}'_{2}(\tilde{{\mathcal{J}}}_{23}-\tilde{{\mathcal{J}}}_{32}),\\
\Lambda_{3}^{\rm{em}}&=&
\mathcal{J}'_{2}{\mathcal{J}}_{pp}-\mathcal{J}'_{3}({\mathcal{J}}_{23}-
{\mathcal{J}}_{32}), \quad
\tilde{\Lambda}_{3}^{\rm{em}}=\mathcal{J}'_{2}\tilde{{\mathcal{J}}}_{pp}-
\mathcal{J}'_{3}(\tilde{{\mathcal{J}}}_{23}-\tilde{{\mathcal{J}}}_{32}).\nonumber
\end{eqnarray}
Again, as in the case of gravitational radiation, we use values for
infinite series (\ref{l:sum}) to obtain the total electromagnetic
radiation rates of energy, momentum and angular momentum:
\begin{equation}\label{Pe:fin}
  \frac{d\dot{P}^{\mu}_{\rm em}}{d\Omega}
  =n^{\mu}\frac{q^{2}\mu}{32\pi^{3}}\int
  d^{4}\xi\mathcal{P}^{\rm{em}}(\Delta
  x\,\rm{mod}\,2\pi-\pi)^{2},
\end{equation}
\begin{eqnarray}
  \label{Le:fin} \frac{d\dot{L}_{v}^{\rm{em}}}{d\Omega}&=&
  \frac{\mathcal{L}q^{2}\mu}{128\pi^{4}}\int
  d^{4}\xi\left[\frac{1}{3}\left((\Delta x\,\rm{mod}\,2\pi-\pi)^{3}- \pi^{2}
  \Delta x \,\rm{mod}\,2\pi\right)
  \tilde{\Lambda}_{2}^{\rm{em}}\right.\nonumber\\
  &&\left. -(\Delta
  x\,\rm{mod}\,2\pi-\pi)^{2}\Lambda_{2}^{\rm{em}}\right],\nonumber\\
  \frac{d\dot{L}_{w}^{\rm{em}}}{d\Omega}&=&
  -\frac{\mathcal{L}q^{2}\mu}{128\pi^{4}}\int
  d^{4}\xi\left[\frac{1}{3}\left((\Delta x\,\rm{mod}\,2\pi-\pi)^{3}- \pi^{2}
  \Delta x \,\rm{mod}\,2\pi\right)
  \tilde{\Lambda}_{3}^{\rm{em}}\right.\nonumber\\
  &&\left. -(\Delta
  x\,\rm{mod}\,2\pi-\pi)^{2}\Lambda_{3}^{\rm{em}}\right].
\end{eqnarray}
As a result we found expressions for electromagnetically radiated energy,
momentum and angular momentum from chiral string loops in which the
summation over modes $l$ is carried out.

\section{Numerical examples}{\label{sec:ex}}

In this Section we apply the derived analytical formulas (\ref{P:fin}),
(\ref{L:fin}), (\ref{Pe:fin}) and (\ref{Le:fin}) for correspondingly
gravitational and electromagnetic radiation to some particular examples
of chiral loops. On the final steps the numerical calculations of
four-dimensional integrals are used to find the energy, momentum and
angular momentum radiation rates as the functions of current on the
string.

We first consider the class of the piece-wise linear kinky loops. Let
$\mathbf{a}(\xi)$ and $\mathbf{b}(\eta)$ be piece-wise linear functions,
i.~e. vector functions $\mathbf{a}(\xi)$ and $\mathbf{b}(\eta)$ are closed
loops, consisted of the connected straight parts. The join points of
segments of $a-$ and $b-$ loops, where $\mathbf{a}'(\xi)$ and
$\mathbf{b}'(\eta)$ discontinuous, called ``kinks''. We take the $a$
-loop consisting of $N_{a}$ and $b$ -loop consisting of $N_{b}$ segments
(parts). Kinks are numbered by indexes $i=0,1,..,N_{a}-1$, the value of
$\xi$ on the kink, numbered by $i$ we note as $\xi^{i}$. In the following
we will denote the numbers of segments by up-indexes and tensor components
by lower indexes respectively.  As we will use only space tensor
components there would be no confusion. Without the loss of generality we
can set $\xi^{0}=0$. Note that $\xi^{i+N_{a}}=\xi^{i}+2\pi$ due to
periodicity. Denoting further $\Delta\xi^{i}=\xi^{i+1}-\xi^{i},\,
\mathbf{A}^{i}=a(\xi^{i}),\,
\mathbf{a}^{i}=(\mathbf{A}^{i+1}-\mathbf{A}^{i})/{\Delta\xi^{i}}$, and
similar denoting for $b-$ loop, we find
\begin{eqnarray}
\label{2-ab1}
\mathbf{a}(\xi)&=&\mathbf{A}^{i}+(\xi-\xi^{i})\mathbf{a}^{i},\,
\xi\in[\xi^{i},\xi^{i+1}],\nonumber\\
\mathbf{b}(\eta)&=&\mathbf{B}^{j}+(\eta-\eta^{j})\mathbf{b}^{j},\,
\eta\in[\eta^{j},\eta^{j+1}].
\end{eqnarray}
In the case of piece-wise linear loops the functions $\mathcal{I}_{p}$,
$\mathcal{Y}_{p}$, $\mathcal{M}_{pq}$, $\mathcal{N}_{pq}$, $\mathcal{X}$
and $\mathcal{Z}_{p}$ in (\ref{IYMN2}) and (\ref{XZ2}) become the sums of
delta-functions because of the discontinuity of $\mathbf{a'}$ and
$\mathbf{b'}$ on the kinks. For example, for function $\mathcal{I}_{p}$
from (\ref{IYMN2}) we have:
\begin{equation}
\label{I:ex}
  \mathcal{I}_{p}=
  \sum_{i}\left(\frac{\mathbf{a}^{i}\mathbf{e_{p}}}{1+\mathbf{a}^{i}\mathbf{n}}
  -\frac{\mathbf{a}^{i-1}\mathbf{e_{p}}}{1+
  \mathbf{a}^{i-1}\mathbf{n}}\right)\delta(\xi-\xi^{i}).
\end{equation}
For others functions one may obtain the similar expressions. Due to the
presence of delta-functions in $\mathcal{I}_{p}$, $\mathcal{Y}_{p}$,
$\mathcal{M}_{pq}$, $\mathcal{N}_{pq}$, $\mathcal{X}$ and
$\mathcal{Z}_{p}$ the integrations in (\ref{P:fin}), (\ref{L:fin}),
(\ref{Pe:fin}) and (\ref{Le:fin}) could be easily carried out. To obtain
the expressions for gravitational and electromagnetic radiation from
general formulas one should replace integrations in (\ref{P:fin}),
(\ref{L:fin}), (\ref{Pe:fin}) and (\ref{Le:fin}) by summations over the
kinks and make the following substitutions:
\begin{eqnarray}
\label{sub}\Delta x&\rightarrow&x^{ijkl}=
  \xi^{i}-\xi^{k}-(\eta^{j}-\eta^{l})+
  \mathbf{n}(\mathbf{a}^{i}-\mathbf{a}^{k}+\mathbf{b}^{j}
  -\mathbf{b}^{l}),\nonumber\\
  \mathcal{I}_{p}&\rightarrow&
  \mathcal{I}^{i}_{p}=\frac{\mathbf{a}^{i}\mathbf{e_{p}}}{1
  +\mathbf{a}^{i}\mathbf{n}}-  \frac{\mathbf{a}^{i-1}\mathbf{e_{p}}}{1
  +\mathbf{a}^{i-1}\mathbf{n}}, \\
  \mathcal{Y}_{p}&\rightarrow& \mathcal{Y}^{j}_{p}=
  \frac{\mathbf{b}^{j}\mathbf{e_{p}}}{1-\mathbf{b}^{j}\mathbf{n}}-
  \frac{\mathbf{b}^{j-1}\mathbf{e_{p}}}{1-\mathbf{b}^{j-1}\mathbf{n}}.
  \nonumber
\end{eqnarray}
The similar substitutions should be performed for functions $M_{pq}$,
$N_{pq}$, $X$ and $Z_{p}$.

\subsection{2-2 piece-wise loop}

\begin{figure}[t]
\includegraphics[angle=-90, width=340 pt]{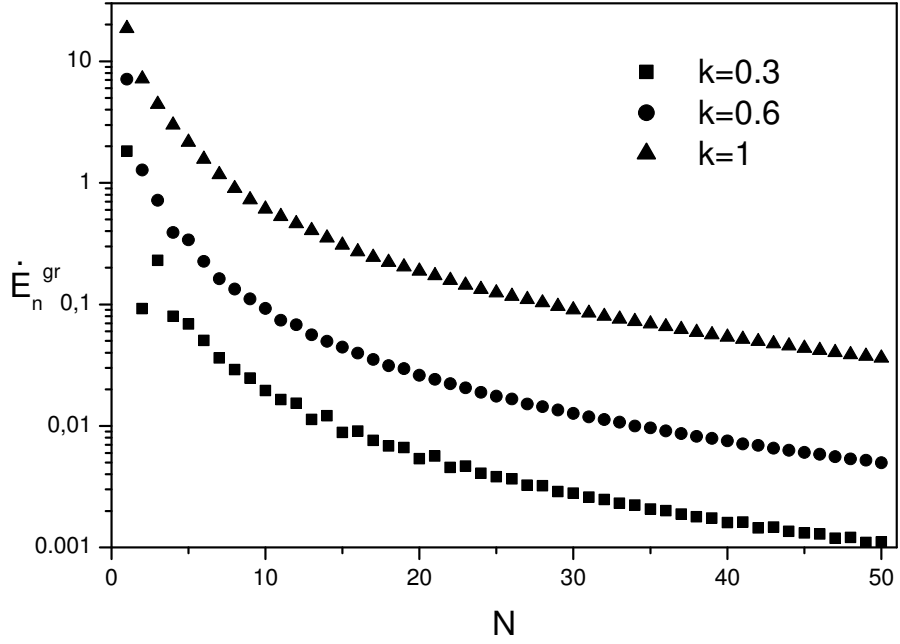}
\caption{\label{ENG2-2} Radiated gravitational energy, $\dot{E}_{n}^{gr}$
in units $G\mu^{2}$. For the 2-2 kinky loop the energy radiation is drawn
logarithmically as a function of the mode number $N$ in units of
$G\mu^{2}$ for different values of parameter $k$.}
\end{figure}

\begin{figure}[t]
\includegraphics[angle=-90, width=340 pt]{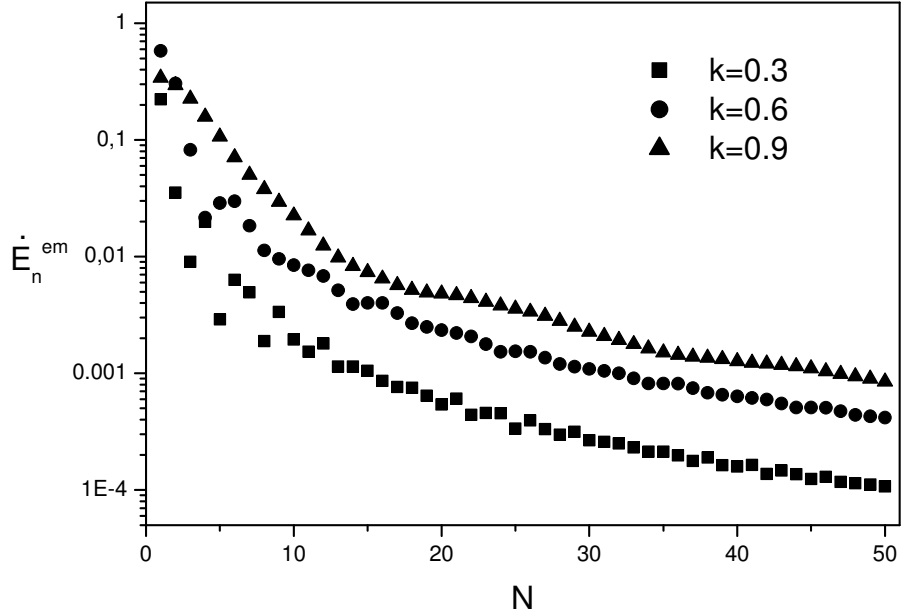}
\caption{\label{ENEm2-2}Radiated electromagnetic energy,
$\dot{E}_{n}^{em}$ in units $q^{2}\mu$. For the 2-2 kinky loop the energy
radiation is drawn logarithmically as a function of the mode number $N$
in units of $q^{2}\mu$ for different values of parameter $k$.}
\end{figure}

\begin{figure}[t]
\includegraphics[angle=-90, width=340 pt]{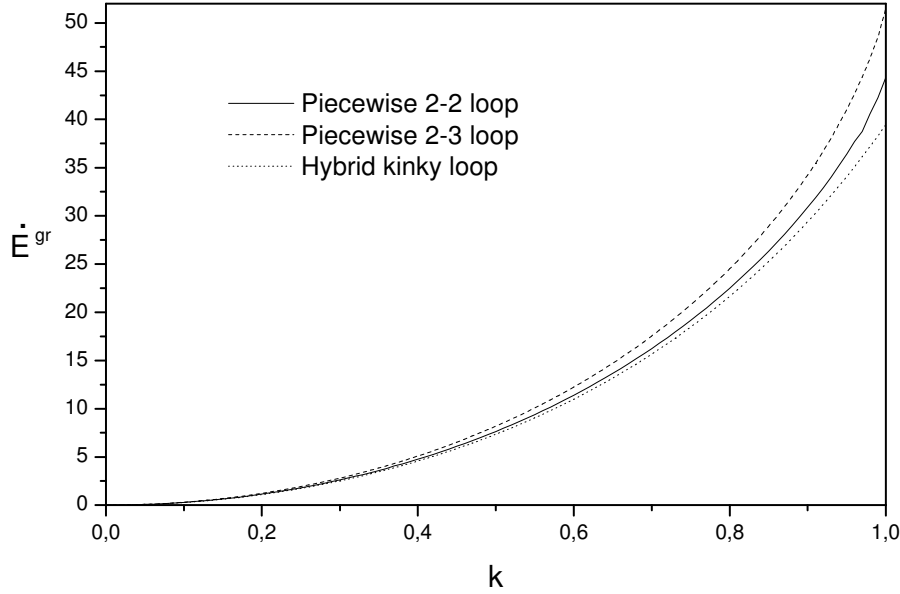}
\caption{\label{EGCL2322}Total radiated gravitational energy
$\dot{E}^{gr}$ in units $G\mu^{2}$ for the 2-2, 2-3 piecewise and hybrid
kinky loops is shown as a function of parameter $k$. For 2-2 loop
$\alpha=\pi/2$, for 2-3 loop $\beta=\pi/4$, for hybrid loop $\gamma=0$.}
\end{figure}

\begin{figure}[t]
\includegraphics[angle=-90, width=340 pt]{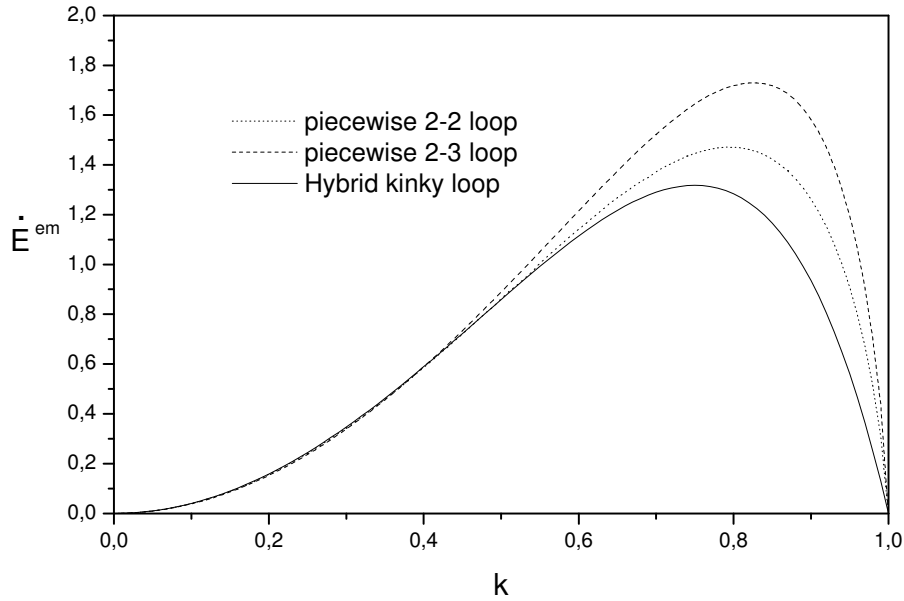}
\caption{\label{EEmCL2322}Total radiated electromagnetic energy
$\dot{E}^{em}$ in units $q^{2}\mu$ for the 2-2, 2-3 piece-wise and hybrid
kinky loop is shown as a function of parameter $k$. For 2-2 loop
$\alpha=\pi/2$, for 2-3 loop $\alpha=\pi/4$, for hybrid loop $\alpha=0$.}
\end{figure}

As the first example let us consider the following chiral string loop:
\begin{equation}
  \label{ex1}
  \mathbf{a}\!=\!\mathbf{A}\!\left\{\!
    \begin{array}{lcl}\left(\xi-\pi/2\right), &&\!\!\! 0\le\xi\le\pi, \\
  \left(3\pi/2-\xi\right), &&\!\!\!   \pi\le\xi\le 2\pi, \\
    \end{array} \right.
  \mathbf{b}\!=\! k\mathbf{B}\!\left\{\!
    \begin{array}{lcl}\left(\eta-\pi/2\right), &&\!\!\! 0\le\eta\le\pi,\\
  \left(3\pi/2-\eta\right),&&\!\!\!  \pi\le\eta\le 2\pi, \\
    \end{array} \right.
\end{equation}
where $\mathbf{A}$ and $\mathbf{B}$ are arbitrary constant unit vectors
with the angle $\alpha$ between them. This is the generalization of
ordinary piece-wise linear loop without the current \cite{Garfinkle1} to
the chiral current case. The both $a$- and $b$-loops consist of two linear
segments. We call this loop as 2-2 piece-wise loop. It has such a
symmetry that no momentum or angular momentum are radiated. The
dependence of radiated electromagnetic and gravitational energy on the
mode number $l$ is shown in Fig.~\ref{ENG2-2} and \ref{ENEm2-2} for the
case of 2-2 piece-wise loop with $\alpha=\pi/2$. The decreasing of
radiated energy with mode number $l$ is more pronounced for larger
current, as it should be physically, because the maximal speed of the
string is decreasing with the increasing of the current. In the case of
electromagnetic radiation besides the overall decreasing of energy with
the mode number one can see in Fig.~\ref{ENEm2-2} also the weak
oscillations of energy rate. The period of this oscillation is increasing
with $k$. In the Fig.~\ref{EGCL2322} and \ref{EEmCL2322} the dependence of
total radiated energy is shown as a function of parameter $k$ for
$\alpha=\pi/2$. It is seen the monotonous increasing of the gravitational
energy radiation with the $k$ (i.~e. with the decreasing of the string
current). At the same time the electromagnetic energy radiated by the
string has a maximum near $k\sim 0.9$. For $k=0$ (no current on the
string) our result for gravitational radiation
$\dot{E}^{\rm{gr}}_{\rm{max}}=44.36 G\mu^2$ reproduces the result of
Garfinkle and Vachaspati \cite{Garfinkle1}. The corresponding graphs for
other values of $\alpha$ look similar. The maximum values of energy rates
$\dot{E}^{\rm{gr}}_{\rm{max}}$ and $\dot{E}^{\rm{em}}_{\rm{max}}$
increase with decreasing of $\alpha$. While $k$ at which the maximum is
reached for gravitational radiation rate is $k=1$ exactly and for
electromagnetic radiation rate $k\sim 0.9$ respectively.

\subsection{2-3 piece-wise loop}

\begin{figure}
\includegraphics[angle=-90, width=340 pt]{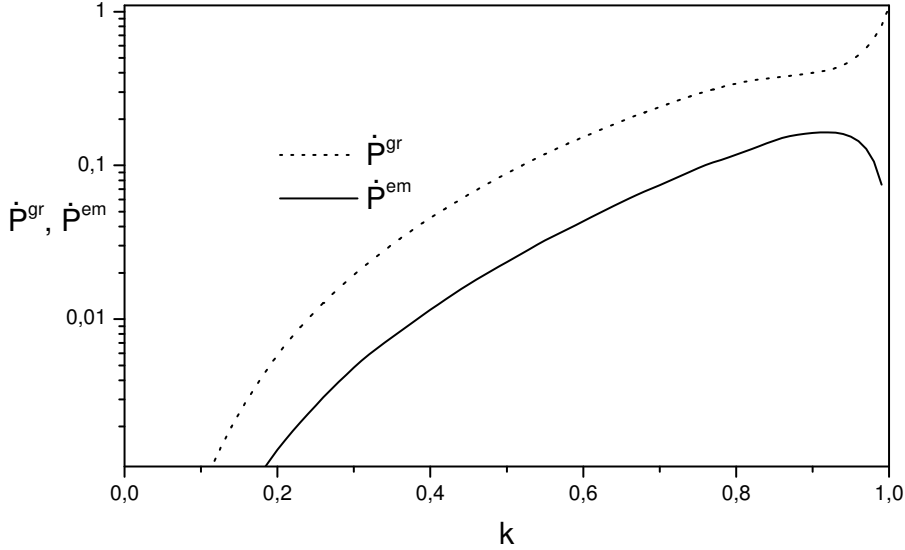}
\caption{\label{PGEm2-3}Total radiated electromagnetic $|\dot{P}^{em}|$
momentum in units $q^{2}\mu$ and gravitational $|\dot{P}^{gr}|$ momentum
in units $G\mu^{2}$ for 2-3 kinky loop is shown as a function of
parameter $k$.}
\end{figure}

\begin{figure}
\includegraphics[angle=-90, width=340 pt]{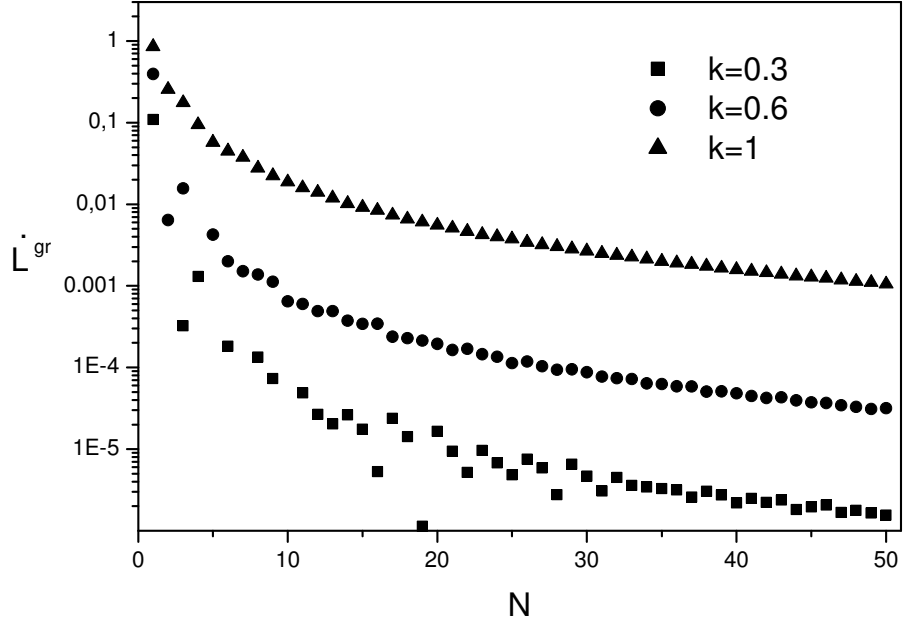}
\caption{\label{LNG2-3}Radiated gravitational angular momentum
$\dot{L}^{gr}$ in units $\mathcal L G\mu^{2}$ for the 2-3 kinky loop is
shown as a function of mode number $N$ for different values of parameter
$k$.}
\end{figure}

\begin{figure}
\includegraphics[angle=-90, width=340 pt]{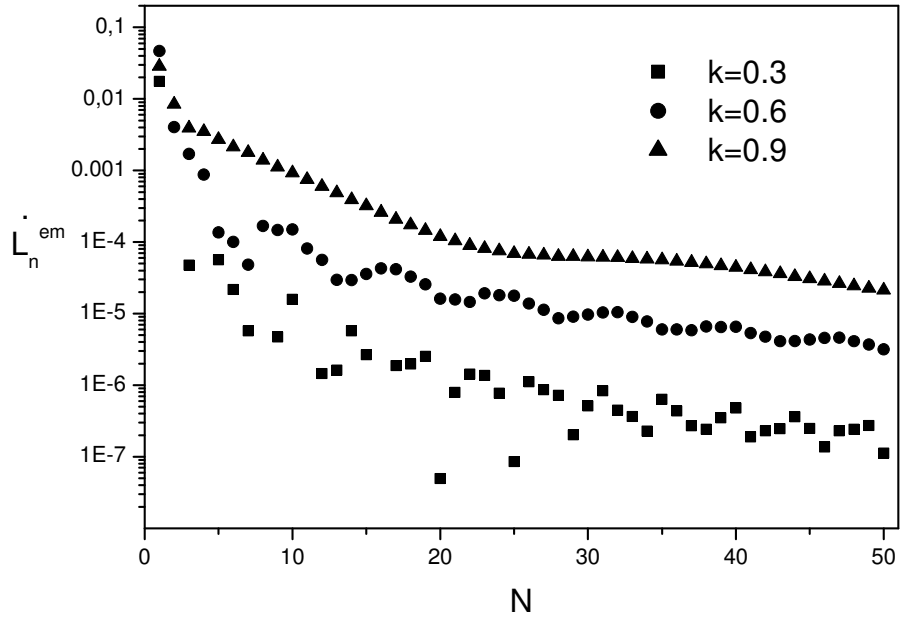}
\caption{\label{LNEm2-3}Radiated electromagnetic angular momentum
$\dot{L}^{em}$ in units $\mathcal L q^{2}\mu$ for the 2-3 kinky loop is
shown as a function of mode number $N$ for different values of parameter
$k$.}
\end{figure}

\begin{figure}
\includegraphics[angle=-90, width=340 pt]{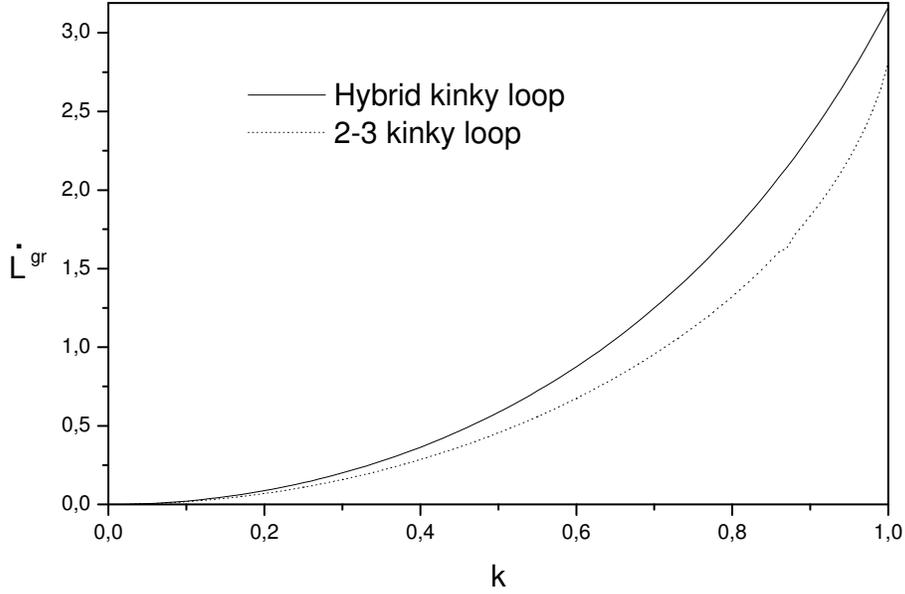}
\caption{\label{LGCL2-3}Total radiated gravitational angular momentum
$\dot{L}^{gr}$ in units $\mathcal L G\mu^{2}$ for 2-3 kinky loop and
hybrid loop is shown as a function of parameter $k$.}
\end{figure}

\begin{figure}
\includegraphics[angle=-90, width=340 pt]{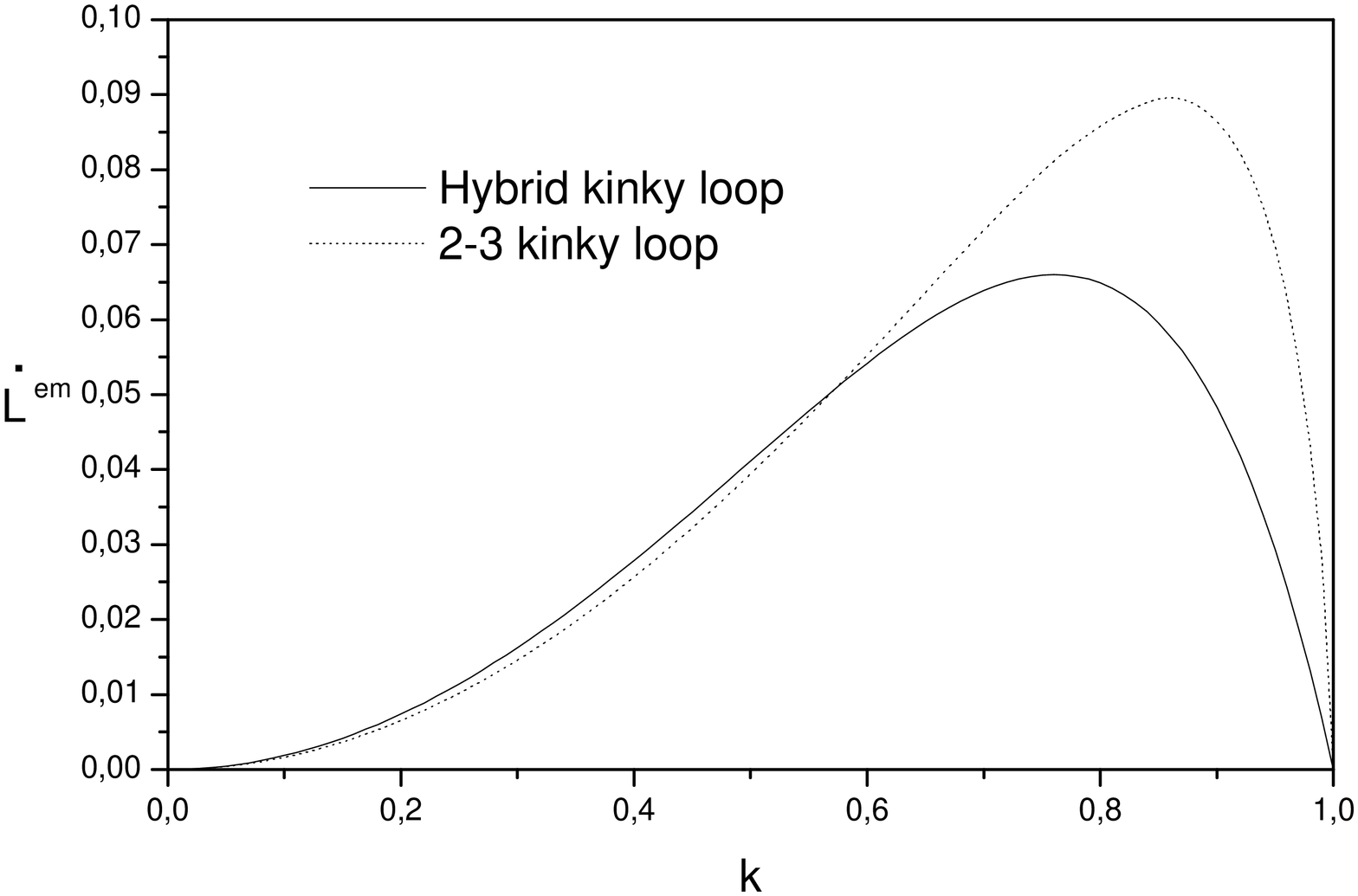}
\caption{\label{LEmCL2-3}Total radiated electromagnetic angular momentum
$\dot{L}^{em}$ in units $\mathcal L q^{2}\mu$ for 2-3 kinky loop and
hybrid loop is shown as a function of parameter $k$.}
\end{figure}

As the second example we consider the piece-wise linear loop of the
following configuration (the generalization of the loop, considered in
\cite{Allen3}): $a$-loop consists of 2 segments and lies along the $z$-
axis. One kink of $a$-loop is positioned at the origin ($\xi=0$) and the
another kink ($\xi=\pi/2$) has coordinates $(0,0,\pi/2)$. The positions of
kinks of $b$-loop are given by the following coordinates: the first kink
at $\eta=0$ is positioned at the origin; the second kink at $\eta=\pi/3$
has coordinates $-({k\pi}/{3})(\cos\beta,\sqrt{3},\sin\beta)$ and the
third kink at $\eta=2\pi/3$ has coordinates
$({k\pi}/{3})(\cos\beta,-\sqrt{3},\sin\beta)$. We call this loop as 2-3
piece-wise loop. Total radiated energy rates into the gravitational and
electromagnetic waves for $\beta=\pi/4$ are shown in Fig.~\ref{EGCL2322}
and \ref{EEmCL2322}. This loop radiates also momentum and angular
momentum. In the Fig.~\ref{PGEm2-3} the total momentum rates into
electromagnetic and gravitational waves are shown for $\beta=\pi/4$. The
corresponding rates for angular momentum as a function of the mode number
are shown in Fig.~\ref{LNG2-3} and Fig.~\ref{LNEm2-3} for $\beta=0$. Again
one can see that the angular momentum radiation rate into the
electromagnetic waves weakly oscillates with mode number. The total
angular momentum radiations are shown in Fig.~\ref{LGCL2-3} and
Fig.~\ref{LEmCL2-3}. The graphs for momentum and angular momentum rates
look very similar to graphs for energy radiation: in the case of
gravitational radiation the increasing of corresponding rates are
monotonous with $k$. While for electromagnetic radiation the radiation of
momenta has maximums near $k=0.9$.

\subsection{Hybrid kinky loop}

As the third example let us consider the loop of the following
configuration:
\begin{equation}
  \label{ex:3}
  \mathbf{a} = (\sin\xi,\; -\cos\xi,\; 0),\quad
  \mathbf{b} = k \mathbf{B}\left\{
    \begin{array}{lcl}\left(\eta-\pi/2\right), & &
  0\le\eta\le\pi,\\
  \left(-\eta+3\pi/2\right), & &   \pi\le\eta\le 2\pi. \\
    \end{array} \right.
\end{equation}
The $a$-loop in this example is a circle in the $(x,y)$ plane and
$\mathbf{B}\!=\!(\cos\gamma;0;\sin\gamma)$. For $\gamma=\pi/2$ the
gravitational and electromagnetic radiated energy and angular momentum
 are shown on the Fig.~\ref{EGCL2322}, \ref{EEmCL2322}, \ref{LGCL2-3} and
\ref{LEmCL2-3}. The total gravitational energy radiation for $k=1$
coincides with the result of Allen et.~al \cite{Allen2}
($\dot{E}^{\rm{gr}}\simeq 39.0 G\mu^{2}$). For $\gamma=\pi/4$ the
results are presented in the Table~\ref{Table1}.
\begin{table}[t]
\caption{The radiation rates from hybrid kinky loop} \label{Table1}
\begin{tabular}{|c|c|c|c|c|c|}
\hline
k & 0.00 & 0.25 & 0.50 & 0.75 & 1.00\\
\hline
$\dot{E}^{\rm gr}$ & 0.00 & 1.79 & 7.56 & 19.10 & 46.58\\
$\dot{E}^{\rm em}$ & 0.00 & 0.23 & 0.86 & 1.60 & 0.00\\
$\dot{L}^{\rm gr}$ & 0.00 & 0.19 & 0.81 & 1.11 & 6.20\\
$\dot{L}^{\rm em}$ & 0.00 & 0.008 & 0.029 & 0.059 & 0.0\\
\hline
\end{tabular}
\end{table}

\begin{table}[b]
\caption{The cosine of deviation angle between
$\dot{\mathbf{L}}^{\rm{gr}}$ and $\dot{\mathbf{L}}^{\rm{em}}$ and
$\mathbf{L}_{st}$}\label{Table2}
\begin{tabular}{|c|c|c|c|c|c|}
\hline &k&0.25&0.5&0.75&1\\
\hline 2-3 loop,&$\varepsilon^{gr}$&-0.94 &-0.94 &-0.95 &-0.96 \\
\cline{2-6}$\beta=\pi/4$&$\varepsilon^{em}$&-0.98 &-0.98
&-0.99&-\\
\hline hybrid loop,&$\varepsilon^{gr}$&-0.97 &-0.97 &-0.97&-0.97\\
\cline{2-6}$\gamma=\pi/4$&$\varepsilon^{em}$&-0.78 &-0.88
&-0.98&-\\
\hline
\end{tabular}
\end{table}

Durrer in \cite{Durrer} found that the radiated angular momentum for some
particular class of ordinary cosmic string loops $\mathbf{\dot{L}}^{gr}$
is antiparallel to the stationary angular momentum $\mathbf{L}_{st}$ of
the loop. It means that angular momentum of the loops always decreases
with time due to gravitational radiation. Our results for angular momentum
radiation into electromagnetic and gravitational waves for string loops
with chiral current agree in general with the results of Durrer. The
chiral loops considered in this paper also lose angular momentum with
time. But unlike the examples considered by Durrer we found that for some
configurations of chiral loops $\mathbf{\dot{L}}^{\rm gr}$ and
$\mathbf{\dot{L}}^{\rm em}$ are not exactly antiparallel to total angular
momentum of the loop $\mathbf{L}_{\rm st}$, but deviate on some small
angle. In the Table~\ref{Table2} the values $\varepsilon^{\rm
gr}=(\mathbf{\dot{L}}^{\rm gr}\mathbf{L}_{\rm st})/|\mathbf{\dot{L}}^{\rm
gr}| |\mathbf{L}_{\rm st}|$ and $\varepsilon^{\rm
em}=(\mathbf{\dot{L}}^{\rm em} \mathbf{L}_{\rm st})/|\mathbf{\dot{L}}^{\rm
em}| |\mathbf{L}_{\rm st}|$, determining the angle between
$\mathbf{\dot{L}}$ and $\mathbf{L}_{\rm st}$ are presented for 2-3
piece-wise loop with $\beta=\pi/4$ and hybrid kinky loop with
$\gamma=\pi/4$.  Note that for symmetric configurations $\beta=0$ and
$\gamma=0$ the angular momentum radiation $\mathbf{\dot{L}}^{\rm gr}$,
$\mathbf{\dot{L}}^{\rm em}$ is exactly antiparallel to $\mathbf{L}_{\rm
st}$ at any $k$.

\section{Conclusion} \label{sec:co}

We found the general formulas for gravitational and electromagnetic
energy, momentum and angular momentum radiation rates into the unit solid
angle from chiral cosmic string loops. The main new result of the paper
is the presentation of the radiation rates from oscillating string loops
in the integral form. In the corresponding integrals the summations of
infinite mode series have been performed analytically. The derived
expressions for $d \dot{E}/d\Omega$, $d\dot{\mathbf{P}}/d\Omega$ and
$d\dot{\mathbf{L}}/d\Omega$ contain four-dimensional integrals, which
depend on the particular loop configuration. The derived integral
presentation is especially convenient for numerical calculations in
comparison with a weakly convergent summation over modes. To find the
total rates of radiated energy, momentum and angular momentum one should
integrate the obtained expressions over unit sphere. The final
expressions for gravitational and electromagnetic radiation can be
written in the following form:
\begin{eqnarray}
\label{gamma1} \dot{E}^{\rm{gr}}=\Gamma_{E}^{\rm{gr}} G\mu^{2},\,
\dot{P}^{\rm{gr}}=\Gamma_{P}^{\rm{gr}} G\mu^{2},\,
\dot{L}^{\rm{gr}}=\Gamma_{L}^{\rm{gr}} \mathcal L G\mu^{2},\nonumber\\
\dot{E}^{\rm{em}}=\Gamma_{E}^{\rm{em}} \mu q^{2},
\dot{P}^{\rm{em}}=\Gamma_{P}^{\rm{em}} \mu q^{2},
\dot{L}^{\rm{em}}=\Gamma_{L}^{\rm{em}} \mathcal L \mu q^{2},
\end{eqnarray}
where numerical coefficients $\Gamma$ depend on the loop configuration
(and also on the current along the loop). Applying our formulas to some
examples of chiral string loop configurations we calculated numerically
coefficients $\Gamma$ as functions of $k$.

In this paper the following three family of examples have been considered:
(i) the piece-wise linear kinky loop with $a$ and $b$-loop consisting of
two straight parts (2-2 piece-wise loop); (ii) the piece-wise linear loop
such that $a$-loop consists of two segments and $b$-loop consists of three
segments (2-3 piece-wise loop); (iii) the hybrid loop in which $a$-loop is
circle and $b$-loop consists of two straight parts (hybrid kinky loop).
For first and second examples the four-dimensional integrals in our
expressions for radiated energy, momentum and angular momentum become the
multiple sums over the kinks. These sums can be analytically calculated
using the symbolic computation on computer (e.~g. ``Mathematica''
packet). To obtain the radiation for third example (hybrid loop) we
calculated two-dimensional integrals (originated from the smooth
$a$-loop) and summed over the kinks of $b$-loop. Unfortunately, we could
not carry out the calculations for strings with $a$ and $b$ loops being
arbitrary smooth closed curves because the corresponding calculations of
four-dimensional integrals take an enormous amount of time.

For considered examples we observe weak oscillations of the
electromagnetic radiation as a function of mode number. These
oscillations (accompanied with a general decreasing of the radiation rate
with mode number) have the different periods depending on the current
along the string: the larger current the smaller the period of
oscillations. This effect does not take place for gravitational radiation.

The total gravitational radiation of energy, momentum and angular
momentum behave in a similar way. They increase slowly with $k$, when $k$
is small (and the current is large) and rapidly increase at large $k$ (or
at large current). In total the gravitational radiation rates are
increasing monotonous functions of $k$. For the electromagnetic radiation
the situation is different: the losses of energy, momentum and angular
momentum into electromagnetic waves for all considered examples have
maximum near $k\sim 0.9$, i.~e. when the current is rather small. For
considered examples the maximal coefficients in (\ref{gamma1}) have the
following values:
\begin{eqnarray}
\label{gamma2} \Gamma_{E}^{\rm{gr}}&\simeq& 50, \quad
\Gamma_{P}^{\rm{gr}}\simeq 1, \quad \Gamma_{L}^{\rm{gr}}\simeq 3,
\nonumber\\
\Gamma_{E}^{\rm{em}}&\simeq& 2, \quad \Gamma_{P}^{\rm{em}}\simeq
0.1, \quad \Gamma_{L}^{\rm{em}}\simeq 0.1.
\end{eqnarray}

We also have found, that for some non-symmetric examples of chiral loops,
the radiated angular momentum $\dot{\mathbf{L}}$ into electromagnetic and
gravitational waves is not exactly opposite to the angular momentum of
the loop $\mathbf{L}_{st}$, but slightly differ from it (even when there
is no current on the string), unlike the other types of loops considered
by Durrer \cite{Durrer}.

\newpage

\end{document}